\documentstyle[12pt]{article}
\begin{document}
\thispagestyle{empty}
\begin{center}
{\LARGE\bf International Workshop on the Future of Physics and Society}\\[2 mm] 
{\large Debrecen, Hungary, 4--6 March, 1999}\\[15 mm]
{\Large \bf Workshop Summary}\\[3 cm]
Raymond S. Mackintosh\\
Physics Department, The Open University\\
Milton Keynes, MK7 6AA, UK

\end{center}

\vskip 1.5cm
{\bf Abstract}\,\, The Debrecen workshop was one of a number held in preparation
for the UNESCO--ICSU World Conference on Science, which will take place
in Budapest, June, 1999. A report representing the views
of the workshop, prepared for that conference and containing a number 
of recommended actions,
is included with this summary. The workshop affirmed the 
ongoing importance of physics for its own sake and as part of our culture,
as a key element in increasingly unified science  and as an essential contributor
to the solution of environmental and energy problems. The problems
faced by physics as an activity and as an educational subject were discussed and
actions for both society as a whole and the physics community itself were put forward.

\vfill

\today

\newpage
\begin{center}
{\Large\bf Introduction}
\end{center}
A principal function of the workshop was to submit a report
making recommendations to the UNESCO--ICSU World Conference on Science, to be
held in Budapest, June, 1999. Nevertheless, a great many important
points were raised which are 
addressed to the international physics community rather than to 
the World Conference. 

This workshop summary is therefore in two parts: the first
part is exactly the report finally submitted to UNESCO and the second
is a summary of the other points raised at the conference which were 
agreed to be important.

\vskip 1.5cm
\begin{flushleft}
{\Large\bf  Part I: Report to the World Conference on Science}\\[5 mm]
\end{flushleft}
\begin{center} 
{\large\bf Preface}
\end{center}

The workshop affirmed three general conclusions:
\begin{enumerate}
\item The contribution of physics to all aspects of life, material and
non-material, will be essential for the foreseeable future.
\item Physics currently faces serious problems in the world. Many of these
problems affect science in general, but a number are specific to physics. 
\item Actions are needed to assure the continued health of 
physics research, teaching and cultural influence.
Some form of `contract' between physicists and the rest of 
society will be required.

\end{enumerate}
 
We emphasise that the problems physics faces are not related to the 
subject matter but to its relations with society and the perceptions of
society. By `physics' we include the physical sciences in general and we affirm
the growth of interdisciplinary fields and the trend for areas such as 
astronomy, cosmology, environmental studies and biophysics 
to become ever more closely linked
with all aspects of physics.

The workshop identified  seven important actions. 
 We recommend that the World Conference on 
Science organised by UNESCO and ICSU consider these for inclusion
in its report ``Science Agenda -- 
Framework of Actions".  Many of them apply to other branches of science and hence
`physics' could be replaced in many places by `science'. Some
actions, however, are specific to physics.
 
The list of recommended actions follows.
We present in appendices some of the points which led us to make
these recommendations. The workshop expressed the view that the experience of
the many relevant professional bodies should be exploited in implementing the
recommendations.

\begin{center}{\large\bf Recommended Actions}\end{center}
\begin{enumerate}
\item Promulgate a declaration affirming the vital importance of 
{\bf basic physical science} and the need to
protect and support  curiosity-led physics.
\item Affirm the  importance of making a substantial effort 
to educate and inform the {\bf public}.
A guideline should be 
established recommending that, say, 1\,\% of money spent on research  
should be made available for public awareness. 
\item Provide substantial support for the improvement of the {\bf teaching} 
of physics throughout the world, at all
levels from school to university. This should involve: 
\begin{itemize}
\item  establishing guidelines for what level of scientific understanding would be expected
at particular  stages of school education and how much time should be  
devoted to physics teaching at each level;
\item monitoring these standards and defending them from external threat;
\item encouraging both curricula and teaching methods to adapt 
to the changing social and scientific environment. 
\end{itemize}
In addition, support is required for {\bf teachers}, for example by enhancing their 
prestige and providing continuing education and personal development. 
UNESCO should promulgate the principle
that {\bf physics should be taught by  persons who have been trained to 
become physics teachers.}  
Reliable {\bf information concerning curricula in different countries} 
should be established and made widely accessible. 
\item Explore ways of establishing a recognised authoritative and {\bf impartial 
international body}, set up under the auspices of UN or UNESCO, 
to adjudicate damaging disputes involving
scientific issues. Examples of such disputes are cold fusion and
a wide range of environmental issues.
The new body would investigate the extent to which claims 
are based upon  established science or are simply
ungrounded opinion, perhaps influenced by pressure groups. 
This will provide an authoritative scientific basis for important political
decisions. 
\item Establish means for supporting physics within the {\bf new democracies} of Europe.
This should be done by facilitating international
collaboration and by encouraging the  support of physicists within their own countries.
 Find ways to support and utilise for mutual benefit the reservoir 
of advanced expertise in the  former Soviet Union.
\item Special measures should be taken to ensure the {\bf free movement} of
scientists. In particular, UNESCO should encourage governments
to facilitate the issuing of visas for scientists if such are required.
\item The long-term health of physics requires the establishment of
{\bf guidelines linking 
R\&D expenditure to GNP} at a level appropriate to the economic state of each
country. In addition, there should be guidelines and standards for coherent and
stable national science policies; these policies should be developed in close
consultation with national scientific communities. UNESCO should establish 
a committee to make recommendations to governments.
\end{enumerate} 
In addition,  the workshop agreed that there are a number of measures which
the physics community itself should take. These will be publicised in due course.

\newpage
\begin{center}{\large\bf Appendix 1: Why the contribution of physics will
continue to be essential}\end{center}
\begin{enumerate}
\item Physics is a central part of our culture and will continue to inspire many people.
Physics  reveals important universal truths notwithstanding certain
strands of postmodern thought.
\item Physics will continue to underpin all science and technology for the foreseeable
future.
\item Physics is and will continue to be {\em essential\/} for analysing and solving 
urgent environmental and energy problems.
\item  Physics plays a unique educational r\^ole: \begin{description}
   \item[Secondary school:] It is recognised that other scientific disciplines more 
                and more require knowledge of physics. 
   \item[Undergraduate:] Physics is becoming recognised as providing education of 
                great value for many careers outside physics such as commerce, 
 		banking and medicine. 
   \item[Doctoral:] PhDs who go into industry are an {\em indispensable\/} byproduct
		of {\em pure physics\/} research.
   \end{description}
\item Physics is {\em global\/} and constitutes our best `anti-Babel'. Generations of
physicists of the most diverse political and cultural backgrounds have collaborated
on the basis of  shared understanding and shared ideals.
\item Physics  sets standards of rational thought in the face of irrationality; 
it upholds the primacy of observation. 	 
\end{enumerate}

\vskip 1cm
\begin{center}{\large\bf Appendix 2: Some general problems currently faced by 
science}\end{center}
\begin{enumerate}
\item Many people feel that science robs the world of meaning and this deeply
affects their attitude to science.
Science is felt by many people to be `cold' and `alienating'.
\item Modern forms of irrationality are becoming widespread and sometimes
involve outright opposition to scientific attitudes and even scientific
knowledge. There is sometimes an unfortunate, even dangerous, political aspect.
\item There is a serious `authority problem' in modern life with few people
able to make rational judgements as to who or what to believe. This is
reflected in a widespread relativism  improperly invoking Einstein. Similarly, 
Heisenberg is improperly invoked in promoting the idea that everything is uncertain anyway. 
The widespread tendency to adopt conspiracy theories  is a potentially
dangerous aspect of this problem. There is a corresponding tendency in academe 
in the form  of social constructivism; 
in extreme form this denies that science can progressively approach universal truth.
\item External pressures, sometimes commercial in nature and often exacerbated
by funding problems, lead to damaging conflicts  within subject areas. Damaging conflicts
also arise between subject areas, particularly under pressure of inadequate funding.
\item In Europe and other places there is a squeeze on industrial research as a result
of `short-termism'.
\item Science teaching and research face specific local problems, particularly
in Eastern Europe and elsewhere. 
\end{enumerate}
It is precisely the  nature of many of these items which makes greater support for
science an urgent matter in the modern world. The workshop also identified a series of
problems specific to physics and a report discussing these as well as some proposed
measures will be published.

\vskip 1.5cm
\begin{center}
{\Large\bf  Part II: Physics in the modern world:\\[2 mm]
some problems and possible solutions}\\[5 mm]
\end{center}
The impact of the globalization on all our institutions and our  value
systems was a common element in many contributions. It is clear that 
physics will have a key
role to play in studying and solving the global environmental
and energy problems the world will face in the coming century.
 Globalization was felt in another way:
while some of the problems listed below are particular to
specific regions,  there was nevertheless very much common ground
in the identification of the general problems faced.

\begin{center} 
{\large\bf 1. Some problems we face}
\end{center}

The workshop identified many difficulties faced by physics as an 
`institution' and as a subject in schools and universities.
These difficulties do not arise from its own subject matter
and in particular the conference affirmed that the subject is  certainly not 
`worked out.' Nevertheless, physics as an activity and as an academic
subject does face problems and some of the specific points 
raised were:

\begin{enumerate}
\item For many students, physics can seem remote from their everyday concerns.
This is true also  for the general public.
This is in great measure because physics is abstract and lacks
visualizable elements (particularly  modern microscopic physics, 
with astrophysics an exception). This presents a problem
for teachers and those communicating with the public. 
\item The fact that physics is essentially mathematical also presents special
problems. While the mathematical language is a main strength of physics 
as a discipline, it is a major obstacle in the way of communicating the meaning of 
physics to the general public.
\item Many school science curriculums are relatively static and remote from
exciting contemporary developments and unrelated to important contemporary
issues such as medicine, energy and the environment. This is in spite of the
direct relevance of physics to all these issues.
\item Physicists have acquired a negative image in some parts of society, 
not least because of the association with nuclear weapons. 
\item The public has no clear picture of how society has benefited from physics
and how physics is essential for solving environmental and energy problems.
\item There is no `physics industry' in the sense that, for example, 
there is a `chemical industry' and a `biotechnology industry'. The
following two problems are, in part, consequences of this.
\item Students in schools are unaware of the career possibilities enabled by
education in physics which exist even in countries in which
high-technology industry is not strong.
\item Physics faces problems in universities: in many places there are
fewer students, and many appear to be less able. Sometimes,
multi-disciplinary courses at undergraduate level add to the 
downward trend in the academic level of courses. This lowering
of standards also occurs as  a result of pressure
to `satisfy customers'. The supply of students to do PhDs is highly susceptible
to economic circumstances and many countries frequently face a serious
shortage. 
\item In Europe and other places there is a squeeze on industrial research as a
result of `short-termism'.
\item In many countries there is a squeeze on {\em pure\/} research  and a
growing requirement for researchers to justify their work in terms of
economic benefits.
\item  In many countries there is a serious lack of competent and
enthusiastic physics teachers.  
\item Physics is particularly subject to competition from pseudo-science.
 This is an aspect of the authority problem:
the public is confused as it is confronted with a mixture of 
information and misinformation through the media, including the Internet.
\end{enumerate}

\vskip 1cm
\begin{center}
{\large\bf 2. Hopeful factors we should find ways to exploit}
\end{center}
The workshop discussed solutions to these problems and also identified some hopeful
signs. 
 Among the positive points were:
\begin{enumerate}
\item Politicians at the highest levels are beginning to find 
that the prestige arising from national
success in pure science is of value in international negotiations. A related
fact is that, in many countries, it is success in science (along with sport)
which most arouses national pride.
\item In some countries, and potentially everywhere, there is a higher 
than ever interest in popular science. This point has been emphasized 
by professionals in the popular science business, and is also clear 
from the number of books published. (The simultaneously
existing problems remind us that the `public' is not a single 
undifferentiated body.)
\end{enumerate}
Our defence of physics, as well as science in general, must
find  ways of exploiting these  hopeful points.  It was pointed out that 
`The resource of the 21st century is knowledge\ldots' and certainly physical
knowledge will be an important part of this.

\begin{center}
{\large\bf 3.  Recommendations to the physics community}
\end{center}

The workshop identified a number of areas where action by the
physics community and its friends, including those involved in teaching physics,
could be of great benefit:
 
\begin{enumerate}
\item Physicists should present a united front; suppress factional fighting; 
show
respect for different subject areas. (We are vulnerable to `divide and rule'.) 
\item Physicists must deal responsibly with the public, avoid exaggeration,
{\em be honest\/} and  should not infringe conventions relating to peer review
and publication. (`Going public' prior to peer review has been very damaging
to biology, and physics has also been harmed by it.)
\item Physicists should assume more responsibility in the issues of the
global environment, sustainable growth or equilibrium and the energy problem.
Physics will have a key role to play in finding an acceptable 
solution to these problems. Particular presentations to the workshop
made very clear  the seriousness of the situation and exemplified
the contribution of physics. 
\item Facilitate improved means for scientists to advise (and enter into
dialogue with) government and other public organisations. (Interaction should 
be both ways and involve the grass roots scientists.)
\item We should find ways of using the expertise of sociologists to explore
in greater depth the cause and nature or anti-scientific feeling; this could
even lead to  {\em entente\/} between physics and 
some part, at least, of the world of sociology. 
This could be of great benefit. An urgent problem requiring study is
 the way the  media treat pseudo-science in modern pluralistic societies.
\item We should find ways to encourage industry to support long term and 
curiosity-led research. Governments should be persuaded to encourage, 
facilitate or enforce this (through tax laws, etc.).
\item Research should be carried out, with the participation of both scientists
and eco\-nom\-ists, which shows the {\em long term\/}
influence of scientific research on GNP. This should be done in a way which
includes such things as the contribution of the training which is an important
byproduct of pure research at PhD level.  
\item Many points relating to {\em teaching\/} physics were mentioned, and some
appear in the `action' statements to UNESCO. Particular points are:
\begin{itemize} \item Physics teaching must respond to changing {\em social\/}
and also {\em scientific} circumstances.
\item There is much value in courses which relate the important findings and
perspectives of cosmology etc.\ to common human needs and aspirations. This was
demonstrated to the workshop by an account of a general course at undergraduate level.
\item Teachers should recognise the value of relating physics teaching to
matters of everyday importance, including environmental and energy issues. 
Teachers should emphasise that it is everybody's moral duty to have an 
elementary understanding of the physics of the threatened global environment. 
The abstract aspects of physics should be moderated at the introductory level. 
\item There are many `modern physics' topics which can be made very accessible
with imaginative teaching methods involving pupil activity. A case was put that
they can be made more accessible and more relevant than some traditional
topics if they are presented with appropriate explanations.
\end{itemize}
Evidently there is a need for continuing debate concerning the teaching of
physics in schools. There is no accepted general solution to the apparently
contradictory requirements of, on the one
hand, attracting talented young people into physics and preparing them for
university level studies, and, on the other hand, teaching physics in a way
that does not repel and alienate  future citizens. 
\item Various points were put forward concerning means to educate and inform the 
public, (the subject of a recommended UNESCO action). Points mentioned include:
the need to professionalize interaction with the media; the need for humour; 
demonstrating the openness of science by letting scientific disputes be
public; the virtue of science laboratories, travelling exhibitions,
science\&technology weeks; the importance of the personal and biographical
elements in presentations, etc. 
\item Investigate and seek remedies for the anomalously low women's participation in
physics in some countries compared to others. We should do this 
in the first place because of the human fulfilment
and beneficial productivity which is currently being lost. There is further
potential benefit:  the remedy may substantially
improve the public status of physics in general.
\end{enumerate}

\begin{center}{\bf Acknowledgements}\end{center}
I am deeply grateful to Herwig Schopper and Rezs\H{o} Lovas for their thorough 
critique of this summary and for saving me from embarrassing omissions.
The first section, the report for UNESCO, was a joint submission of
the three of us on behalf of the workshop. 
The conference was supported by the UNESCO-Physics Action Council, the
European Physical Society, OMFB, OTKA,
MTA and MAL\'EV. I am personally very grateful to the Lovases for hospitality.

\end{document}